\documentclass[pre,11pt,superscriptaddress,preprint]{revtex4-1}

\usepackage{bm,soul}
\usepackage{latexsym}
\usepackage{graphics}
\usepackage{graphicx} 
\usepackage{amsthm} 
\usepackage{amsmath}
\usepackage{wasysym}
\usepackage{amssymb}
\usepackage{enumerate}
\usepackage{slashed}
\usepackage{empheq}
\usepackage[dvips]{epsfig}
\usepackage[colorlinks = true]{hyper ref}
\usepackage{float}  
\usepackage{multirow}
\usepackage{cancel}
\usepackage{bm}
\usepackage[table]{xcolor}
\newcommand{\beq}{\begin{equation}}
\newcommand{\eeq}{\end{equation}}
\newcommand{\nn}{\nonumber \\}
\newcommand{\bl}[1]{{\color{blue}#1}}  
\def\bea{\begin{eqnarray}}
\def\eea{\end{eqnarray}}
\hypersetup{
    bookmarks=true,         
    unicode=false,          
    pdftoolbar=true,        
    pdfmenubar=true,        
    pdffitwindow=false,     
    pdfstartview={FitH},    
    pdftitle={My title},    
    pdfauthor={Author},     
    pdfsubject={Subject},   
    pdfcreator={Creator},   
    pdfproducer={Producer}, 
    pdfkeywords={keyword1} {key2} {key3}, 
    pdfnewwindow=true,      
    colorlinks=true,       
    linkcolor=magenta, 
    citecolor=blue,        
    filecolor=magenta,      
    urlcolor=cyan           
} 

\begin{document}
\title{Entanglement entropy of the large $N$ Wilson-Fisher conformal field theory}
\author{Seth Whitsitt}
\affiliation{Department of Physics, Harvard University, Cambridge MA 02138, USA}   
\author{William Witczak-Krempa}
\affiliation{D\'epartement de Physique, Universit\'e de Montr\'eal, Montr\'eal (Qu\'ebec), H3C 3J7, Canada}
\affiliation{Department of Physics, Harvard University, Cambridge MA 02138, USA}   
\author{Subir Sachdev}
\affiliation{Department of Physics, Harvard University, Cambridge MA 02138, USA}   
\affiliation{Perimeter Institute for Theoretical Physics, Waterloo, Ontario, Canada N2L 2Y5}

\date{\today
\\
\vspace{0.4in}}

\begin{abstract}
We compute the entanglement entropy of the Wilson-Fisher conformal field theory (CFT) in 2+1 dimensions 
with O($N$) symmetry in the limit of large $N$ for general entanglement geometries. We show that the leading
large $N$ result can be obtained from the entanglement entropy of $N$ Gaussian scalar fields with their mass determined
by the geometry. For a few geometries, the universal part of the entanglement entropy of the Wilson-Fisher CFT equals that of a CFT of $N$ massless scalar fields. However, in most cases, these CFTs have a distinct universal entanglement entropy even at $N=\infty$.
Notably, for a semi-infinite cylindrical region it scales as $N^0$ in the Wilson-Fisher theory, in stark contrast
to the $N$-linear result of the Gaussian fixed point. 
\end{abstract}

\maketitle   

\section{Introduction}

The entanglement entropy (EE) has emerged as an important tool in characterizing strongly 
interacting quantum systems \cite{CC04,RT06,Kitaev06,Levin06,Fradkin06,Dong08,Hsu09,MFS09,CH12,Zhang-Grover12}.        
In the context of relativistic theories in 2 spatial dimensions, the so-called $F$ theorem uses the 
EE on a circular disk to place constraints on allowed 
renormalization group flows \cite{RMAS10,RMAS11,JKPS11,KPS11,KPSS12,CH12,TG14}. 
For quantum systems with holographic duals, the EE can be computed via the Ryu-Takayanagi 
formula \cite{RT06}, and this is a valuable tool in restricting possible holographic duals of strongly interacting theories \cite{OTU12,HSS12}. 

Despite its importance, the list of results for the EE of strongly interacting gapless field 
theories in 2+1 dimensions is sparse. 
The most extensive results are for CFTs on a circular disk geometry in the vector 
large-$N$ and small-$\epsilon$ expansions \cite{KPS11,KPSS12,Giombi14,Fei15,Giombi15,Fei16,Tarnopolsky16}.      
Some results have also been obtained \cite{MFS09}
in the infinite cylinder geometry in an expansion in $\epsilon=(3-d)$, where $d$ is the spatial dimension, but the extrapolation of these results
to $d=2$ is not straightforward.   

In this paper we show how the vector large $N$ expansion can be used to obtain the EE in essentially all entanglement
geometries, generalizing results that were only available so far in the circular disk geometry. The large $N$ expansion was also used
in Ref.~\cite{MFS09} in the infinite cylinder geometry, but the results were limited to the universal deviation of the EE when the CFT is tuned away from the critical point by a relevant operator. For a region with a smooth boundary, the groundstate
of a CFT has an EE $S$ which obeys
\beq \label{defgamma}
S = C \frac{L}{\delta} - \gamma 
\eeq
where $\delta$ is a short-distance UV length scale, $C$ is the area law coefficient depending on the regulator, 
$L$ is an infrared length scale associated with the entangling geometry,  
and $\gamma$ is the universal part of the EE we are interested in. We will compute $\gamma$ for the Wilson-Fisher CFT with O($N$)
symmetry on arbitrary smooth regions in the plane, and 
in the cylinder and torus geometries. Our methods generalize to other geometries, and also to other CFTs with a vector 
large $N$ limit. We also obtain universal entanglement entropies associated with geometries with sharp corners.

Our analysis relies on a general result which will be established in Section~\ref{sec:mapping}. 
We consider the large $N$ limit of the Wilson-Fisher CFT on a general geometry using the replica method, which requires the
determination of the partition function on a space which is a $n$-sheeted Riemann surface. The large $N$ limit maps the CFT to 
a Gaussian field theory with a self-consistent, spatially dependent mass \cite{MFS09}. Determining this mass for general $n$
is a problem of great complexity, given the singular and non-translationally invariant $n$-sheeted geometry; complete results for such
a spatially dependent mass are not available. However, we shall show that a key simplification occurs in the limit $n \rightarrow 1$
required for the computation of the EE: the spatially dependent part of the mass does not influence the value of the EE. This simplification
leads to the main results of our paper. We note here that the simplification does not extend to 
the R\'enyi entropies $n \neq 1$,
so we shall not obtain any results for the R\'enyi entropies of the Wilson-Fisher CFT in the large $N$ limit. 

Section~\ref{sec:mapping} 
will compute the EE for the Wilson-Fisher CFT on arbitrary smooth regions in an infinite plane, 
and for regions containing a sharp corner, in which case (\ref{defgamma}) is modified. In both these cases, 
and for other entangling regions in the infinite plane, the EE is equal
to that of a CFT of $N$ free scalar fields. 
Section~\ref{sec:cylent} will consider the case of an entanglement cut on an infinite cylinder. A non-zero limit of $\gamma/N$ as $N\rightarrow \infty$ was 
obtained in Ref.~\cite{MFS09} for the free field case. We will show that a very different result applies to the Wilson-Fisher
CFT, with $\gamma/N = \mathcal{O}(1/N)$.  Section~\ref{sec:torus} considers the case of a torus with two cuts: here $\gamma/N$ is 
non-zero for both the free field and Wilson-Fisher cases, but the values are distinct from each other.

\section{Mapping to a Gaussian theory}
\label{sec:mapping} 

In this section we consider the EE of the critical O($N$) model at large-$N$, and show that it can be mapped to the EE of a Gaussian scalar field.

\subsection{Replica method}   

We first recall how the EE can be computed in a quantum field theory using the replica method introduced in Refs.~\cite{HLW94,CC04}. The EE associated with a region $A$ is given by
\beq
S = - \mathrm{Tr} \left( \rho_A \log \rho_A \right)
\eeq
where $\rho_A$ is the reduced density matrix in $A$. A closely related measure of the entanglement 
is the $n$-th R\'enyi entropy, which is defined as 
\beq
S_n = \frac{1}{1 - n} \log \mathrm{Tr} \rho_A^n
\eeq
where $n>1$ is an integer. In the replica method, outlined below, the R\'enyi entropies are directly computed from a path integral construction. One can then analytically continue $n$ to non-integer values, and obtain the EE as a limit
\beq
\lim_{n \rightarrow 1} S_n = S
\eeq
Equivalently, one can consider expanding $\log \mathrm{Tr} \rho_A^n$ to leading order in $(n - 1)$, obtaining
\beq
\log \mathrm{Tr} \rho_A^{n} = - (n - 1) S + \mathcal{O}\!\left( (n-1)^2 \right)
\label{smalln}
\eeq
Thus, the small $(n - 1)$ behavior of $\mathrm{Tr} \rho^n_A$ is sufficient to compute the entropy $S$. 

The computation of $\mathrm{Tr} \rho_A^n$ proceeds as follows. We first consider the matrix element of the reduced density matrix between two field configurations on $A$, $\phi'_A(\mathbf x)$ and $\phi''_A(\mathbf x)$. This can be computed using the Euclidean path integral
\beq
\langle \phi'_A(\mathbf{x}) | \rho_A | \phi_A''(\mathbf{x}) \rangle = \mathcal{Z}_1^{-1} \int^{\phi(\mathbf{x} \in A, t_E = 0^+) = \phi''_A(\mathbf{x})} _{\phi(\mathbf{x} \in A, t_E = 0^-) = \phi'_A(\mathbf{x})} \mathcal{D}\phi(\mathbf{x}, t_E) e^{- \mathcal{S}_{E}}
\label{matelem}
\eeq
where $\mathcal S_E$ is the Euclidean action of the system.  
We then write the trace over $\rho_A^n$ in terms of these matrix elements
\beq
\mathrm{Tr} \rho_A^n = \int\! \mathcal{D}\phi_A' \mathcal{D} \phi_A'' \cdots \mathcal{D} \phi_A^{(n)}\;\, \langle \phi'_A | \rho_A | \phi_A'' \rangle \langle \phi''_A | \rho_A | \phi_A''' \rangle \cdots \langle \phi^{(n)}_A | \rho_A | \phi_A' \rangle
\label{cnvlt}
\eeq
Combining Eqns.~(\ref{matelem}) and (\ref{cnvlt}), we obtain the path integral expression for $\mathrm{Tr} \rho_A^n$ as
\beq
\mathrm{Tr} \rho_A^n = \frac{\mathcal{Z}_n}{\mathcal{Z}_1^n}
\eeq
Here, $\mathcal{Z}_n$ is the partition function over the $n$-sheeted Riemann surface obtained by performing the integrations in Eq.~(\ref{cnvlt}). In particular, we consider $n$ copies of our Euclidean field theory, but we glue the spatial 
region $(\mathbf{x} \in A, t_E = 0^+)$ of the $k$th copy to the spatial region $(\mathbf{x}\in A, t_E = 0^-)$ of the $(k+1)$th copy, repeating until we glue the $n$th copy to the first copy. This construction introduces conical singularities at the boundary of $A$.

\subsection{Entanglement entropy for the O($N$) model at large $N$} \label{sec:large-N_EE}
We now specialize to the critical O($N$) model in $(2+1)$-dimensions. We use a non-linear $\sigma$ 
model formulation, writing the $n$-sheeted action as
\bea
\mathcal{S}_n &=& \int d^3 x_n \ \mathcal{L}_n \nn
\mathcal{L}_n &=& \frac{1}{2} \phi_{\alpha} \left( - \partial_n^2 + i \lambda \right) \phi_{\alpha} - \frac{N}{2 g_c} i \lambda
\eea
where $\alpha$ runs from $1$ to $N$ and is summed over. 
Here, $d^3x_n$ and $\partial_n^2$ denote the integration measure and the Laplacian on the $n$-sheeted Riemann surface, respectively. The field $\lambda(x)$ is a Lagrange multiplier enforcing the local constraint $\phi(x)^2 = N/g_c$. In the $N = \infty$ limit, the path integral is evaluated using the saddle point method: 
\bea
\mathcal{Z}_n &=& \int \mathcal{D}\phi \mathcal{D}\lambda \ e^{-\mathcal{S}_n} \nn
&=& \int \mathcal{D}\lambda \ \exp\left[ -\frac{N}{2}\mathrm{Tr} \log \left(-\partial_n^2 + i \lambda \right) + \frac{N}{2 g_c} \int d^3 x \ i \lambda \right] \nn
\Longrightarrow \log \mathcal{Z}_n &=& -\frac{N}{2}\mathrm{Tr} \log \left(-\partial_n^2 + \left\langle i \lambda \right\rangle_n \right) + \frac{N}{2 g_c} \int d^3 x_n \, \left\langle i \lambda \right\rangle_n + \mathcal{O}(1/N) \label{evalZ}
\eea
In the last equality, the saddle point configuration of the field $\lambda$ is determined by solving the gap equation 
\beq
G_n(x,x; \langle i \lambda\rangle_n ) = \frac{1}{N} \langle \phi(x)^2 \rangle_n = \frac{1}{g_c}
\eeq
where $G_n(x,x')$ is the Green's function on the $n$-sheeted surface:
\beq
\left( - \partial_n^2 + \langle i \lambda(x) \rangle_n \right) G_n(x,x'; \langle i \lambda\rangle_n ) = \delta^{3}(x - x' ) \label{eqGn}
\eeq
and the critical coupling is determined by demanding that the gap vanishes for the infinite volume theory on the plane:
\beq
\frac{1}{g_c} = \int \frac{d^{3} p}{\left( 2 \pi \right)^{3}} \frac{1}{p^2}
\eeq

In the absence of the entangling cut, $n=1$, we denote the saddle point value of $\lambda$ as
\beq
\left\langle i \lambda \right\rangle_1 = m_1^2
\eeq
We assume that the one-sheeted geometry is translation-invariant, so $m_1$ is independent of position. On the infinite plane we have $m_1 = 0$, but we will also consider geometries where one or both dimensions are finite,
in which case $m_1$ becomes a universal function of the geometry of the system determined by
\beq
G_1(x,x; m_1^2) = \frac{1}{g_c} 
\eeq

On the $n$-sheeted Riemann surface, $\langle i \lambda(x) \rangle_n$ is always a nontrivial function of position, and the exact form of this function depends on the shape of the entangling surface and the number of Riemann sheets $n$. The problem of determining this function can be addressed numerically for fixed $n$, but for the purposes of obtaining the EE, we only need its spatial dependence to first-order in $(n - 1)$. In particular, we assume that we can write
\beq
\langle i \lambda(x) \rangle_{n} \approx m_1^2 + (n - 1) f(x)
\eeq
for some function of space-time $f(x)$. Then to leading order in $N$ and $(n-1)$, we have
\bea
-\log \mathcal{Z}_n &\approx& \frac{N}{2}\mathrm{Tr} \log \left(-\partial_n^2 + m^2_1 \right) - \frac{N}{2 g_c} \int d^3 x_n \ m_1^2 \nn
&+& \left( n - 1 \right) \frac{N}{2}\mathrm{Tr} \left( \frac{f(x)}{-\partial_1^2 + m_1^2} \right) - \left( n - 1 \right) \frac{N}{2 g_c} \int d^3 x \ f(x)
\label{eqeps}
\eea
Then using the definition of $G_1$ and $m_1$,
\beq
\mathrm{Tr} \left( \frac{f(x)}{-\partial_1^2 + m_1^2} \right) = \int d^3x \ G_1(x,x;m_1^2) f(x) = \frac{1}{g_c} \int d^3x f(x)
\eeq
implying that that last line of Eq.~(\ref{eqeps}) vanishes, and $f(x)$ does not contribute to the EE. After using $\int d^3x_n = n \int d^3x$, we can write
\bea
-\log \frac{\mathcal{Z}_n}{\mathcal{Z}_1^n} = \frac{N}{2}\bigg[\mathrm{Tr}  \log \left(-\partial_n^2 + m^2_1 \right) - n \mathrm{Tr} \log \left(-\partial_1^2 + m^2_1 \right) \bigg]
\eea

This final expression is equal to the quantity $-\log \mathrm{Tr} \rho_A^n$ computed for $N$ free scalars with mass $m_1$ and the action
\beq
\mathcal{L}'_n = \phi_{\alpha} \left( -\partial^2_n + m_1^2 \right) \phi_{\alpha}
\eeq
Therefore, the EE of the critical O($N$) model at order $N$ is equal to the EE of $N$ free scalar fields, where the free fields have the same mass gap as the O($N$) model on the physical, one-sheeted surface. 
Similar results will apply to other large-$N$ vector models. For instance, in Appendix~\ref{ap:GN} we
follow very similar steps to show that
the EE of the fermionic Gross-Neveu model maps to that of $N$ free Dirac fermions. The mass of the 
free fermions is determined self-consistently by the spatial geometry of the physical single-sheeted spacetime.

\subsection{Entanglement entropy on the infinite plane}

We first consider the EE when the system is on the infinite plane. In this case, $m_1 = 0$, and the EE associated with a region $A$ is equal to the EE of $N$ massless free scalars in the same region.

One entangling region for which there are known results is the circular disk. 
According to the F-theorem \cite{CH12}, the universal part of the EE for the disk is given by
\beq
\gamma_{\mathrm{disk}} = F \equiv - \log |\mathcal{Z}_{S^3} |
\eeq
Here, $\mathcal{Z}_{S^3}$ is the finite part of the Euclidean partition function on a three-sphere
spacetime. This quantity was computed in Ref.~\cite{KPS11} for massless free scalar fields and for the large-$N$ O($N$) model, and they were found to be equal at order $N$ in agreement with our general result given above. Explicitly, 
\beq
\gamma_{\mathrm{disk}} = \frac{N}{16} \left( 2 \log 2 - 3 \frac{\zeta(3)}{\pi^2} \right)
\eeq
where $\zeta(3) \approx 1.202$. Our results also apply to regions with sharp corners, in which case we can 
make non-trivial checks of our general result, as we now discuss.  

\subsubsection{Entanglement entropy of regions with corners} 
When region $A$ (embedded in the infinite plane) contains a sharp corner of opening angle $\theta$, 
the EE of a CFT (\ref{defgamma}) is modified by a subleading logarithmic correction \cite{Casini_rev,Hirata} 
\begin{align}
  S = C \frac{L}{\delta} - a(\theta) \log(L/\delta) + \dotsb
\end{align}  
where the dimensionless coefficient $a(\theta)\geq 0$ is universal, and encodes non-trivial information
about the quantum system. Since we work in the infinite plane, according to our analysis above, the large-$N$
value of $a(\theta)$ will be the same as for $N$ free scalars, namely
\begin{align} \label{corner_WF}
  a_{\rm WF}(\theta) = N\, a_{\rm free}(\theta) +\mathcal O(N^0)
\end{align}
The non-trivial function $a_{\rm free}(\theta)$ for a single free scalar   
was studied numerically and analytically by a number of authors 
for a wide range of angles \cite{Casini_rev,corner-prl,Elvang15,Laflorencie15,WKB,Helmes16}.    
Interestingly, we can make an analytical verification of the relation (\ref{corner_WF}) in the nearly smooth limit,
by virtue of the following identity that holds for any CFT \cite{corner-prl,Bueno-Myers15,faulkner15}
\begin{align}  \label{smooth-corner} 
  a(\theta\approx \pi) = \frac{\pi^2 C_T}{24}\, (\theta-\pi)^2 + \mathcal O\!\left((\theta-\pi)^4\right)
\end{align}
Here, $C_T$ is a non-negative coefficient determining the groundstate two-point function of the stress tensor $T_{\mu\nu}$:
\begin{align}  
  \langle T_{\mu\nu}(x) T_{\eta\kappa}(0) \rangle = \frac{C_T}{x^6}\, \mathcal I_{\mu\nu,\eta\kappa}(x)
\end{align}
where $\mathcal I_{\mu\nu,\eta\kappa}(x)$ is a dimensionless tensor fixed by conformal symmetry \cite{Osborn94}.
Eq.~(\ref{smooth-corner}) was conjectured \cite{corner-prl} for general CFTs in two spatial dimensions,  
and subsequently proved using non-perturbative CFT methods \cite{faulkner15}. 
Now, $C_T$ is the same at the Wilson-Fisher and Gaussian fixed points \cite{Sachdev93}     
at leading order in $N$:
\begin{align} \label{CT_WF} 
  C_T^{\rm WF} = N C_T^{\rm free} + \mathcal O(N^0)
\end{align}
which, when combined with Eq.~(\ref{smooth-corner}), leads to a non-trivial confirmation of (\ref{corner_WF}) in the nearly 
smooth limit $\theta\approx \pi$. 
(We note that $C_T^{\rm free}=\tfrac{3}{32\pi^2}$ using conventional normalization \cite{Osborn94}.)       

The knowledge of $C_T$ can be used to make a statement about $a(\theta)$ away from the nearly smooth limit
because the existence of the following lower bound \cite{WKB}:  
$a(\theta) \geq C_T \frac{\pi^2}{3} \log\left[ 1/\sin(\theta/2) \right]$,
which follows from the strong subadditivity of the EE, and (\ref{smooth-corner}). 
We see that applying this bound to the large-$N$ Wilson-Fisher fixed point is consistent with our result (\ref{corner_WF}). 

\section{Infinite cylinder}
\label{sec:cylent}

\begin{figure}
\includegraphics[width=8cm]{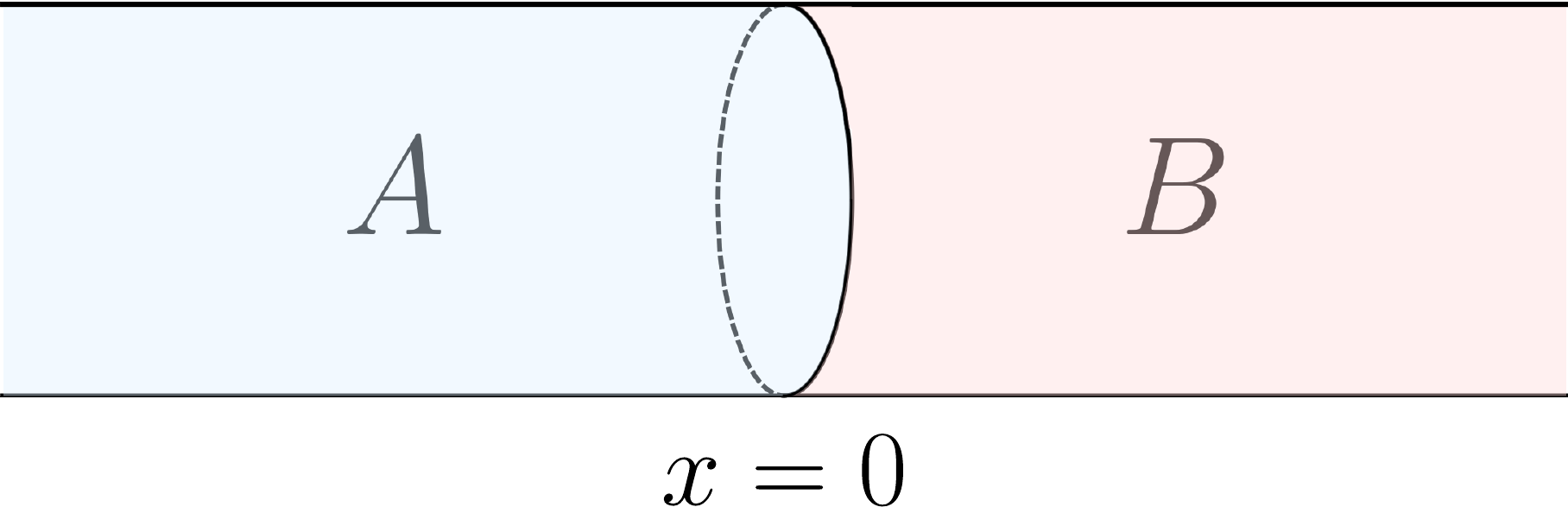}
\caption{The geometry considered in calculating the entanglement entropy on the infinite cylinder.}
\label{cylfig}
\end{figure}

We now compute the EE of the semi-infinite region obtained by tracing out half of an infinite cylinder.  
The relevant geometry is pictured in Fig.~\ref{cylfig}. We can take the position of the cut to be at $x=0$ by translation invariance. 
As for the disk, we can write the EE as 
\beq
S = C \frac{L}{\delta} - \gamma_{\mathrm{cyl}}
\eeq
where $\gamma_{\mathrm{cyl}}$ is the universal part. The existence of a universal $\gamma_{\rm cyl}$ in critical theories was
first established \cite{Hsu09,Oshikawa10,Hsu10} for the $z=2$ quantum Lifshitz model using the methods of Ref.~\cite{Fradkin06}.
In the context of CFTs, this geometry was considered in Ref.~\cite{MFS09},  
where the entropy $\gamma$ was computed for massless free fields and for the  
Wilson-Fisher fixed point in the $\epsilon = (3 - d)$ expansion (where the extra dimensions introduced in the $\epsilon$-expansion are 
made periodic with circumference $L$). 

We first review the calculation of the entropy for free massive fields, which will allow us to calculate the EE for the Wilson-Fisher fixed point for large $N$. We allow for twisted boundary conditions along the finite direction
\beq \label{twist-y}
\phi(x,y + L) = e^{i\varphi_y}\phi(x,y)
\eeq 
Here, $\varphi_y \in [0,2\pi)$. We note that unless $\varphi_y = 0,\pi$, the fields $\phi$ are complex. In this case, we are considering $N/2$ complex fields, and the O($N$) symmetry of the theory breaks down to U(1)$\times$SU($N/2$).

This geometry allows a direct analytic computation of the $n$-sheeted partition function for free fields by mapping to radial coordinates, $(t_E,x) = (r \cos \theta , r \sin \theta)$. In these coordinates, the $n$-sheeted surface is fully parametrized by giving the angular coordinate a periodicity of $2 \pi n$. In Refs.~\cite{CC04,MFS09}, it was shown that the $n$-sheeted partition function for free fields can be written in terms of the one-sheeted Green's function:
\bea
-\log \frac{\mathcal{Z}_n}{\mathcal{Z}_1^n} &=& \frac{N}{2}\bigg[ \mathrm{Tr} \log \left(-\partial_n^2 + m^2 \right) - n \mathrm{Tr} \log \left(-\partial_n^2 + m^2 \right) \bigg] \nn
&=& \frac{\pi N}{6} \left( n - \frac{1}{n} \right) L G_1(x,x;m^2)
\eea
Then using Eq.~(\ref{smalln}), the EE is given by
\beq
S = \frac{\pi N}{3} L G_1(x,x;m^2)
\label{Sfree}
\eeq

In Appendix \ref{app:green}, we compute the Green's function for a massive free 
field on the cylinder. 
Using Eq.~(\ref{gfunc}), and making the cutoff dependence explicit,   
we find the regularized part of the EE to be (see also Ref.~\cite{Chen16}) 
\bea
\gamma_{\mathrm{cyl}} &=& \frac{N}{12} \log \left[ 2 \left( \cosh m L - \cos \varphi_y \right) \right]
\eea
For $m = 0$, this reduces to Eq.~(5.12) of Ref.~\cite{MFS09}, and indeed displays  
a divergence for a periodic boundary condition $\varphi_y=0$ due to the zero mode. 
We note that the universal contribution to the EE of $N/2$ complex free scalar fields is of order $N$, 
as one would expect from a free field theory with $N$ degrees of freedom.

We now turn to the Wilson-Fisher fixed point. In a finite geometry, the Wilson-Fisher fixed point will acquire a mass gap $m_1$ which is proportional to $1/L$ and depends only on $\varphi_y$. This is computed by solving $G_1(x,x;m_1^2) = 1/g_c$, which is done in Appendix \ref{app:green}. The result is
\beq
m_1 = \frac{1}{L} \mathrm{arccosh} \left( \frac{1}{2} + \cos \varphi_y \right)
\label{cylgap}
\eeq
Then from the arguments of Section \ref{sec:mapping}, 
\beq
\gamma_{\mathrm{cyl}} = \frac{N}{12} \log \left[ 2 \left( \cosh m_1 L - \cos \varphi_y \right) \right] = 0
\label{cylchi}
\eeq
It happens that for the saddle point value of the mass, the universal part of the EE vanishes for all values of the twist $\varphi_y$. The leading contribution to $\gamma_{\mathrm{cyl}}$ will be of $\mathcal{O}(N^0)$, a drastic reduction from Gaussian fixed point which is of order $N$.

This result can be seen more directly from Eq.~(\ref{Sfree}). The gap equation implies that $G_1(x,x;m_1^2) = 1/g_c$, so without even solving for $m_1$, the EE can immediately be written 
\beq
S = \frac{\pi N}{3} \frac{L}{g_c}
\eeq
However, the critical coupling is completely non-universal and independent of $L$. Using a hard UV momentum cutoff $1/\delta$,
\beq
\frac{1}{g_c} = \int^{1/\delta} \frac{d^3 p}{\left( 2 \pi \right)^3} \frac{1}{p^2} = \frac{1}{2 \pi^2 \delta}
\eeq
and the EE is pure area law, $S \propto L/\delta$. 

In fact, this result can be extended to other geometries. The result $\gamma_{\mathrm{cyl}} = 0$ for the large-$N$ Wilson-Fisher fixed point occurred because the entropy is proportional to $G_1(x,x;m^2)$. However, the results of Refs.~\cite{CC04,MFS09} imply that the expression for the free-field entropy given in Eq.~(\ref{Sfree}) holds for \emph{any} system where the entangling cut is perpendicular to an infinite, translationally-invariant direction. Thus, if we consider the large-$N$ Wilson-Fisher CFT on any $d$-dimensional spatial geometry with at least one infinite dimension, the universal part of the EE obtained by tracing out over half of that dimension 
is $\mathcal{O}(N^0)$. This argument only holds in dimensions where the Wilson-Fisher CFT exists, so for $1 < d < 3$. In particular, this result agrees with the large-$N$ limit of the $\epsilon$-expansion calculation in Ref.~\cite{MFS09}, which considered the Wilson-Fisher CFT on the $(3 - \epsilon)$-dimensional spatial region $\mathbb{R} \times \mathbb{T}^{2 - \epsilon}$, where $\mathbb T^d$ is the $d$-dimensional torus. 
This constitutes a non-trivial consistency check on both calculations. 

Finally, we note that similar results apply to other large $N$ models. As shown in Appendix \ref{ap:GN}, the EE for the Gross-Neveu CFT maps to that of $N$ free Dirac fermions, where the mass of the fermions is determined by the spatial geometry of the one-sheeted spacetime, $\mathrm{Tr} \ G_1^{F}(x,x;m_1) = m_1/g_c^2$. Here, the critical coupling $1/g_c^2$ is again a non-universal quantity which cannot depend on the spatial geometry of the system, and is proportional to the UV cutoff. Then using the results of Ref.~\cite{CC04}, it can be shown that $S \propto G^F_1$ for free fermions on the spatial geometries discussed in the previous paragraph, and therefore $\gamma = \mathcal{O}(N^0)$ for the large-$N$ Gross-Neveu CFT on the infinite cylinder.

\section{Torus} 
\label{sec:torus}
\begin{figure}
  \centering
  \includegraphics[scale=.5]{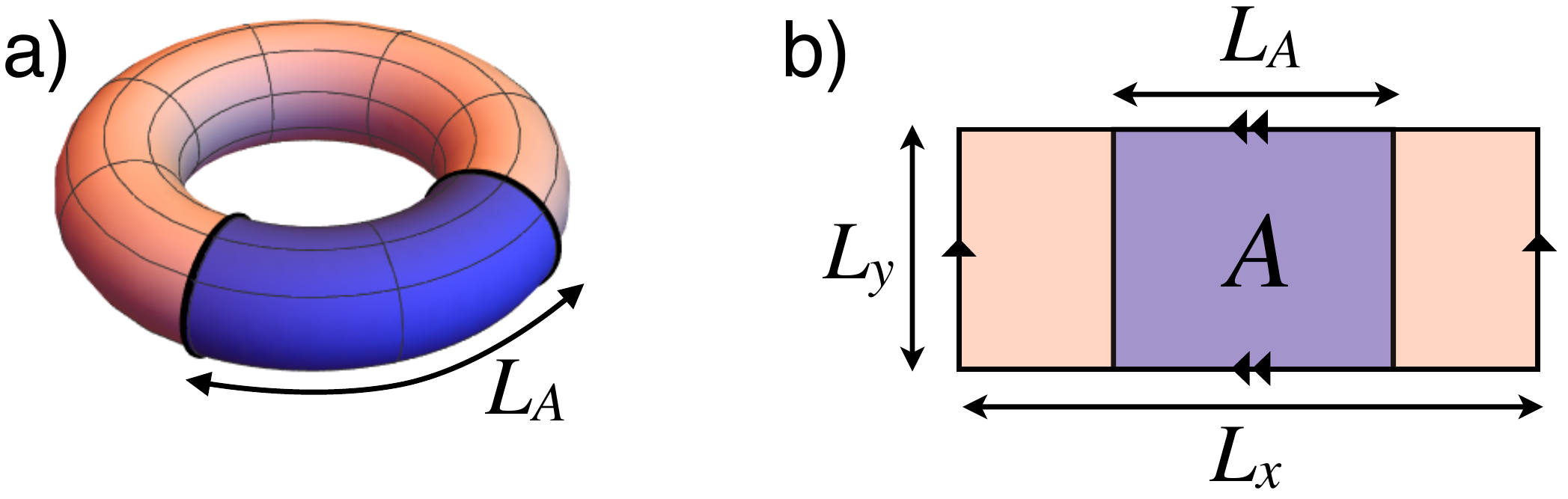} 
  \caption{a) Two dimensional (flat) torus. b) Its representation in the plane. 
We analyze the entanglement entropy of region $A$.} 
  \label{fig:torus2d}
\end{figure} 
We study the EE of the large-$N$ fixed point on a spatial torus, as shown in Fig.~\ref{fig:torus2d}.
\bl{For a CFT}, 
we expect the following form for $S$ \cite{Chen15,Will-torus}
\begin{align} \label{S-torus}
  S= C \frac{2L_y}{\delta} - \gamma_{\rm tor}(u;\tau) 
\end{align} 
where we have defined the ratio  
\begin{align}
  u =  L_A/L_x
\end{align}
and $\tau$ is the modular parameter, $\tau=iL_x/L_y$, for the rectangular torus we work with.
$\gamma_{\rm tor}$ is a universal term that we shall study at the large-$N$ Wilson-Fisher fixed point.

As discussed in Section~\ref{sec:large-N_EE}, the EE at leading order in $N$ is given by that of $N/2$ 
free complex scalars with a mass $m_1$ determined by the geometry.
$m_1$ is thus the self-consistent mass on the torus
for a single copy of the theory, which was recently computed at large $N$ in Ref.~\cite{WS16}. 
It obeys the scaling relation 
$m_1 L_y=g(\tau)$, where $\tau$ is the aspect ratio of the torus, and $g$ is a non-trivial 
dimensionless function given in Appendix~\ref{ap:gap-torus}. $m_1$ depends on both   
twists along the $x$- and $y$-cycles of the torus, $\varphi_x,\varphi_y$.   
Since $\gamma_{\rm tor}^{\rm free}$ for a massive free boson is not known,
we will numerically study the $(u,\tau)$-dependence of $\gamma_{\rm tor}^{\rm WF}$ 
by working on the lattice. 

However, before doing so, we describe two limits in which we can
make statements about $\gamma_{\rm tor}$. First, we consider the so-called thin torus limit 
for which $L_y\to 0$, while $L_A,L_x$ remain finite, i.e.\ $\tau\to i\infty$ and $u$ is fixed.  
For generic boundary conditions, we have that the torus EE approaches twice the 
semi-infinite cylinder value \cite{Chen15,Will-torus} discussed above, $\gamma_{\rm tor}\to 2\gamma_{\rm cyl}$. 
This holds in the absence of zero modes, which is the generic case. Our result Eq.~(\ref{cylchi})
implies that $\gamma_{\rm tor}= \mathcal O(N^0)$ in that limit. However, this cannot hold at all 
values of $\tau$. Indeed, for any fixed $\tau$ let us consider the ``thin slice'' limit $L_A\to 0$.
There, $\gamma_{\rm tor}$ reduces to the universal contribution of a
thin strip of width $L_A$ in the infinite plane \cite{Chen15,Will-torus}, $\gamma_{\rm tor}=\kappa L_y/L_A$,
where the universal constant $\kappa\geq 0$ can be computed in the infinite plane. $\kappa$ is thus independent
of the boundary conditions along $x,y$.
Applying our mapping to free fields, this means that at leading order in $N$
\begin{align} \label{kappa-WF}
  \kappa^{\rm WF}=N \kappa^{\rm free} 
\end{align}
where $\kappa^{\rm free}\simeq 0.0397$ for a free scalar \cite{Casini_rev}. By continuity, we
thus expect that for general $u$ and $\tau$, $\gamma_{\rm tor}^{\rm WF}$ will scale linearly with
$N$ in the large-$N$ limit. We now verify this statement via a direct calculation.

\subsection{Lattice numerics} 
The lattice Hamiltonian for a free boson of mass $m_1$ can be taken to be
\begin{align}
  H &= \frac1 2 \sum_{k_y}\sum_{i=0}^{L_x-1} \left( |\pi_{k_y}(i)|^2 + |\phi_{k_y}(i+1)-\phi_{k_y}(i)|^2+
\omega_{k_y}^2 |\phi_{k_y}(i)|^2 \right) \label{H_latt} \\
  \omega_{k_y}^2 &= 4\sin^2(k_y/2) + m_1^2
\end{align}
where the theory is defined on a square lattice with unit spacing, $\pi_{k_y}(i)$ 
is the operator canonically conjugate to $\phi_{k_y}(i)$, and $|A|^2=A^\dag A$. The index $i$ runs over the $L_x$ 
lattice sites in the $x$-direction.
Crystal momentum along the $y$-direction remains a good quantum
number in the presence of the entanglement cut, and is quantized as follows
\begin{align} \label{ky}
  k_y = \frac{2\pi n_y}{L_y} +\frac{\varphi_y}{L_y}\, , 
\end{align}
where the integer $n_y$ runs from $0$ to $L_y-1$, and $\varphi_y$ is the twist along the $y$-direction. 
We note that the Hamiltonian (\ref{H_latt}) 
corresponds to $L_y$ decoupled 1d massive boson chains: $H=\sum_{k_y} H_{\rm 1d}(k_y)$,
each with an effective mass $\omega_{k_y}$. This means that the EE is the sum over the corresponding 
1d EEs: $S=\sum_{k_y}S_{\rm 1d}(k_y)$. For each 1d chain, the EE for the interval of length $L_A\leq L_x$
is obtained from the correlation
functions $X_{ij}=\langle \phi^\dag(i) \phi(j)\rangle$ and $P_{ij}=\langle \pi^\dag (i) \pi(j)\rangle$,
where we have suppressed the $k_y$ label. The prescription for the EE is then \cite{Casini_rev}
\begin{align}
  S_{\rm 1d} = \sum_\ell \left[ (\nu_\ell +\tfrac1 2)\log(\nu_\ell+\tfrac1 2)- (\nu_\ell -\tfrac1 2)\log(\nu_\ell-\tfrac1 2) \right]
\end{align}
where $\nu_\ell$ are eigenvalues of the matrix $\sqrt{X_AP_A}$, with the $A$ meaning that $X_{ij}$ 
and $P_{ij}$ are restricted to region $A$. This method was previously used to study the EE of free fields on 
the torus \cite{Chen15,Will-torus,PSI_16,Chen16}.   

\begin{figure}
  \centering
  \includegraphics[scale=.5]{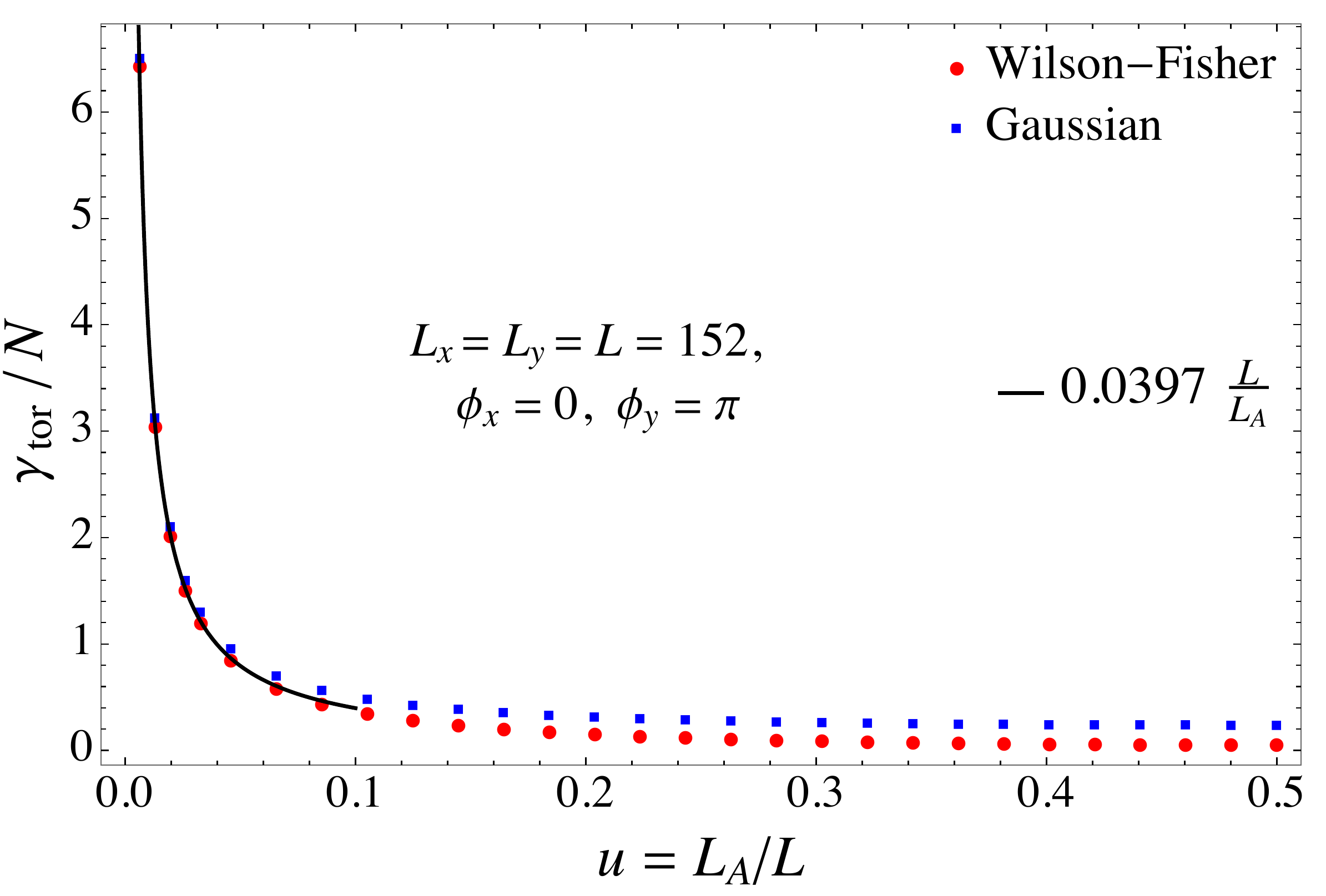} 
  \caption{The universal EE $\gamma_{\rm tor}$ of a cylindrical region, $L_A\times L$, 
cut out of a square torus, $L\times L$.
The red dots correspond to the interacting fixed point of the O$(N)$ model at large $N$, while the blue squares to
the Gaussian fixed point ($N$ free relativistic scalars). We have normalized $\gamma_{\rm tor}$ by $N$. 
The data points were obtained numerically on a square lattice of linear size $L=152$. The line shows the
expected divergence in the small $u$ limit, the same for both theories.}
  \label{fig:torus-WF}
\end{figure}

To obtain the universal part of the entropy we first need to 
numerically determine the area law coefficient $C$ (\ref{S-torus}), which we find is $C\simeq 0.07745$. 
We can then isolate the universal part, $\gamma_{\rm tor}$, by subtracting the area law contribution.
The result for a square torus, i.e.\ $\tau=i$, is shown in Fig.~\ref{fig:torus-WF}, where we compare the Wilson-Fisher fixed point with the Gaussian fixed point. Only $0<u<1/2$ is shown because the other half
is redundant by virtue of the identity $\gamma_{\rm tor}(1-u)=\gamma_{\rm tor}(u)$, true for pure states.
We set $\varphi_x=0$ and $\varphi_y=\pi$ (since fully periodic boundary conditions yield a divergent $\gamma_{\rm tor}^{\rm Gauss}$), 
which leads to a purely imaginary mass $m_1 L_y\simeq i\,1.77078$
for the Wilson-Fisher theory, while $m_1$ is naturally zero at the Gaussian fixed point.
The imaginary mass does not cause a problem since $k_y^2+ m_1^2>0$ in the presence of the twist, (\ref{ky}). 
From Fig.~\ref{fig:torus-WF}, we see that $\gamma_{\rm tor}^{\rm WF}$ scales linearly with $N$ as was anticipated above.
However, contrary to the case of the infinite plane, the EE of the Wilson-Fisher fixed point 
is reduced compared to the Gaussian fixed point, 
$\gamma_{\rm tor}^{\rm WF}(u) < \gamma_{\rm tor}^{\rm Gauss}(u)$ for all values of $u$. The difference between
the EE of the two theories decreases in the thin slice limit $u\to 0$, where we have the divergence 
$\gamma_{\rm tor}=\kappa/u$, with the same constant $\kappa$ for both theories, Eq.~(\ref{kappa-WF}). This constant
has been calculated in a different context \cite{Casini_rev}, $\kappa=N\kappa^{\rm free} =N\,0.0397$, and fits
our data very well. Another consistency check is that $\gamma_{\rm tor}(u)$ should be convex decreasing \cite{Will-torus}
for $0<u<1/2$, which is indeed the case in Fig.~\ref{fig:torus-WF}.

\section{Conclusions}
\label{sec:conc}

The large $N$ limit of the Wilson-Fisher theory yields the simplest tractable strongly-interacting CFT in 2+1 dimensions.
In this paper, we have succeeded in computing the entanglement entropy of this theory using a method which can 
be applied to essentially any entanglement geometry. In particular, for any region in the infinite plane,
the EE of the large-$N$ Wilson-Fisher fixed point is the same as that of $N$ free massless bosons. 
In contrast, when space is compactified into a cylinder or a torus, the results will differ in general
as we have illustrated using cylindrical entangling regions. Our results naturally extend to other large-$N$ vector theories,
like the fermionic Gross-Neveu CFT (Appendix~\ref{ap:GN}).

In the case of the EE of the semi-infinite cylinder, $\gamma_{\rm cyl}$, Ref.~\cite{MFS09} has compared its value
at the UV Gaussian fixed point and the IR Wilson-Fisher one using the $\epsilon$ expansion. 
\bl{For these two specific fixed points,} it was found that $|\gamma_{\rm cyl}^{\rm UV}| > |\gamma_{\rm cyl}^{\rm IR}|$, suggestive of 
 the potential of $\gamma_{\rm cyl}$ to ``count'' degrees of freedom. Our results at large $N$, however,
show that the opposite can \bl{occur for certain RG flows. Namely,} $|\gamma_{\rm cyl}^{\rm UV}| < |\gamma_{\rm cyl}^{\rm IR}|$
\bl{is possible}.
Indeed, let us consider the flow from the Wilson-Fisher \bl{critical} point to the \bl{stable} fixed point \bl{describing the phase where 
the O$(N)$
symmetry is spontaneously broken} \cite{ssbook}. 
\bl{In the UV,} we see that $\gamma_{\rm cyl}^{\rm WF}\sim N^0$ \bl{does not grow with $N$}, while at the IR fixed point $\gamma_{\rm cyl}^{\rm broken}\sim N$
due to the $(N-1)$ Goldstone bosons \cite{Max2011}. This holds for generic boundary conditions $\varphi_y\neq 0$. 

It is also of interest to obtain the R\'enyi entropies of such an interacting CFT, notably for comparison with
large-scale quantum Monte Carlo simulations which can usually only 
yield the second R\'enyi entropy \cite{Hastings10,Inglis_2013}.
Unfortunately, this is a much more
challenging problem, because the full $x$-dependence of the saddle-point $\langle i \lambda(x) \rangle_n$ in (\ref{eqGn}) is needed on a $n$-sheeted
Riemann surface. Numerical analysis will surely be required, supplemented by analytic 
results in the limit of large  and small $x$. \\  

{\bf Note}: While finishing this work, we became aware of a related forthcoming paper \cite{janet-prep} that studies the area law term
in a large-$N$ supersymmetric version of the O$(N)$ model.   

\subsection*{Acknowledgments}

We thank Pablo Bueno, Shubhayu Chatterjee, Xiao Chen, Eduardo Fradkin, Janet Hung, Max Metlitski, and Alex Thomson for useful discussions.
This research was supported by the NSF under Grant DMR-1360789 and the MURI grant W911NF-14-1-0003 from ARO. 
WWK was also supported by a postdoctoral fellowship and a Discovery Grant from NSERC, and by a Canada Research Chair (tier 2). WWK further acknowledges the hospitality of the Aspen Center for Physics, where 
part of this work was done, and which is supported by National Science Foundation grant PHY-1066293.
Research at Perimeter Institute is supported by the Government of Canada through Industry Canada and by the Province of Ontario 
through the Ministry of Research and Innovation. SS also acknowledges support from Cenovus Energy at Perimeter Institute.

\appendix

\section{Green's function and large $N$ mass gap on the cylinder}
\label{app:green}

In Section \ref{sec:cylent}, we used the Green's function for a massive scalar on the infinite cylinder. This is given by
\beq
G_1(x,x;m^2) = \frac{1}{L} \sum_{k_y} \int \frac{d^2 p}{\left(2 \pi\right)^2}  \frac{1}{p^2 + k_y^2 + m^2}
\label{green1}
\eeq
where we allow a twist in the finite direction
\beq
k_y = \frac{2 \pi n_y + \varphi_y}{L}, \qquad n_y \in \mathbb{Z}
\eeq
We evaluate this expression using Zeta function regularization. We first introduce an extra parameter $\nu$, and consider the expression
\beq
G_1(x,x;m^2) = \frac{1}{L} \sum_{k_y} \int \frac{d^2 p}{\left(2 \pi\right)^2}  \frac{1}{\left(p^2 + k_y^2 + m^2\right)^{\nu}}
\eeq 
This expression is convergent for $\nu > 3/2$. We evaluate this expression where it is convergent, and then analytically continue it to $\nu \rightarrow 1$. After evaluating the integrals, we obtain 
\beq
G_1(x,x;m^2) = \frac{1}{8 \pi^2 L \left(\nu - 1 \right)} \sum_{k_y} \frac{1}{\left( k_y^2 + m^2 \right)^{\nu - 1}}
\eeq
The remaining sum needs to be evaluated as a function of $\nu$, which requires the use of generalized Zeta functions. General formulae for sums of this type can be found in Reference \cite{ELI12}, and after evaluating this sum and taking $\nu \rightarrow 1$, we find
\beq
G_1(x,x;m^2) = - \frac{1}{4 \pi L} \log \left[ 2 \left( \cosh m L - \cos \varphi_y \right) \right]
\label{gfunc}
\eeq
We note that the original integral, Eq.~(\ref{green1}), has a linear UV divergence which has been set to zero by our cutoff 
method. In other regularization methods, one generically expects an extra term proportional to the UV cutoff, 
$G_1(x,x;m^2) \propto 1/\delta$, which contributes to the area law in Eq.~(\ref{Sfree}).

We also find the mass gap for the Wilson-Fisher fixed point at large-$N$. The gap equation is
\beq
G_1(x,x;m_1^2) = \frac{1}{g_c}
\eeq
However, in Zeta regularization we have 
\beq
\frac{1}{g_c} = \int \frac{d^3 p}{\left(2 \pi\right)^2} \frac{1}{p^2} = 0
\eeq
Then using Eq.~(\ref{gfunc}), we find the energy gap on the cylinder
\beq
m_1 = \frac{1}{L} \mathrm{arccosh} \left( \frac{1}{2} + \cos \varphi_y \right)
\eeq
as quoted in the main text.

\section{Entanglement entropy of the Gross-Neveu model at large $N$} 
\label{ap:GN} 

We discuss the Gross-Neveu model \cite{zinn2002} in the large $N$ limit. The calculation of the entanglement entropy 
is very similar to the critical O($N$) model, and we find a mapping to the free fermion entanglement analogous 
to the mapping derived in Section \ref{sec:mapping}.  

The critical model is defined by the Euclidean Lagrangian
\beq
\mathcal{L}_n = -\bar{\psi}_{\alpha} \left( \slashed{\partial}_n + \sigma \right)\psi_{\alpha} + \frac{N}{2 g_c^2} \sigma^2
\eeq 
where the repeated index $\alpha$ is summed over, running from 1 to $N$. 
Here, $\sigma(x)$ is a Hubbard-Stratonovich field 
used to decouple the quartic interaction term $(\bar\psi_\alpha\psi_\alpha)^2$.
We now follow the steps in Eq.~(\ref{evalZ}) to obtain the partition function using the saddle point method.
\bea
\log \mathcal{Z}_n = N \ \mathrm{Tr} \log \left( \slashed{\partial}_n + \langle \sigma \rangle_n \right) - \frac{N}{2 g_c^2} \int d^3 x_n \ \langle \sigma \rangle^2_n + \mathcal{O}(1/N)
\eea
The saddle point configuration of $\sigma$ is determined by the Gross-Neveu gap equation
\begin{align}
\mathrm{Tr} \ G^{F}_n(x,x; \langle \sigma\rangle_n ) =& \frac{\langle \sigma \rangle_n}{g_c^2} \nn
\left( \slashed{\partial}_n +  \langle \sigma(x) \rangle_n  \right) G^{F}_n(x,x'; \langle \sigma\rangle_n ) 
=&  \delta^3( x - x' ) 
\end{align}
Here, the trace is over spinor indices and we have left the identity matrix in spinor space implicit. 
The critical coupling is
\beq
\frac{1}{g_c^2} = \left( \mathrm{Tr} \ \mathbb{I} \right) \int \frac{d^{3} p}{\left( 2 \pi \right)^{3}} \frac{1}{p^2}
\eeq

Following our procedure for the O($N$) model, we write the saddle point configuration as
\beq
\langle \sigma(x) \rangle_n \approx m_1 + \left( n - 1 \right) f(x)
\eeq
to leading order in $(n-1)$, for an unknown function $f(x)$. 
Then by a similar reasoning to the calculations in Section \ref{sec:mapping}, we find
\beq
-\log \frac{\mathcal{Z}_n}{\mathcal{Z}_1^n} = -N\bigg[\mathrm{Tr}  \log \left(\slashed{\partial}_n + m_1 \right) - n \mathrm{Tr} \log \left(\slashed{\partial}_1 + m_1 \right) \bigg]
\eeq
This is the $n$-sheeted partition function for $N$ free Dirac fermions with mass $m_1$, where $m_1$ is the mass gap of the Gross-Neveu model on the one-sheeted physical spacetime, ${\rm Tr}\, G^{F}_1(x,x;m_1)=m_1/g_c^2$.

Just as for the O$(N)$ Wilson-Fisher fixed point, we can verify our result for the special case where region $A$
is a disk embedded in the infinite plane. The disk's universal entanglement entropy in the Gross-Neveu CFT
 was found to be that of $N$ free massless Dirac fermions \cite{KPS11},
$\gamma_{\rm disk}= N\gamma_{\rm disk}^{\rm free} + \mathcal O(N^0)$. This is exactly our 
result since $m_1=0$ for this geometry.

\section{Mass gap on the torus} \label{ap:gap-torus} 

In Section \ref{sec:torus}, we used the self-consistent mass of the large-$N$ Wilson-Fisher fixed point on the torus. This mass takes the form 
\beq
L_y m_1 = g(\tau)
\eeq
where $g(\tau)$ is a universal function of them modular parameter of the torus, $\tau$, and the twists $\varphi_x$ and $\varphi_y$. The calculation of $m_1$ was done in Ref.~\cite{WS16}. Here, we outline the results needed for the current work. Unlike Ref.~\cite{WS16}, we specialize to the rectangular torus, $\tau = i L_x/L_y$.

The Green's function on the torus is
\bea
G_1(x,x;m_1^2) &=& \frac{1}{L_x L_y} \sum_{k_x k_y} \int \frac{d \omega}{2 \pi} \frac{1}{\omega^2 + k_x^2 + k_y^2 + m_1^2} \nn
&=& \frac{1}{2 L_x L_y} \sum_{k_x k_y} \frac{1}{\left(k_x^2 + k_y^2 + m_1^2\right)^{1/2}}
\eea
where
\bea
k_x &=& \frac{2 \pi n_x + \varphi_x}{L_x} \nn
k_y &=& \frac{2 \pi n_y + \varphi_y}{L_y}
\eea
for integers $n_x$ and $n_y$. As in Appendix \ref{app:green}, we evaluate $G_1$ using Zeta regularization; the technical details of this calculation can be found in Ref.~\cite{WS16}. In this regularization, the gap equation is
\beq
G_1(x,x;m_1^2) = 0
\eeq

After regularizing, we can write the Green's function as
\bea
4 \pi L_x G_1(x,x;m_1^2) &=& \int_1^{\infty} d\lambda \ \lambda^{-1/2} \exp\left( - \frac{\lambda}{4 \pi}\left( L_y^2 m_1^2 + \frac{\varphi_x^2}{|\tau|^2} + \varphi_y^2 \right) \right) \theta_{3}\left( \frac{i \varphi_x \lambda}{2 \pi |\tau|^2} ; \frac{i \lambda}{|\tau|^2} \right) \theta_{3}\left( \frac{i \varphi_y \lambda}{2 \pi} ; i \lambda \right) \nn
&+ & |\tau| \int_1^{\infty} d\lambda \ \lambda^{-1/2}  \left[\exp\left( - \frac{L_y^2 m_1^2}{4 \pi \lambda} \right) \theta_{3}\bigg( \frac{\varphi_x}{2 \pi} ; i \lambda |\tau|^2 \bigg) \theta_{3}\left( \frac{\varphi_y}{2 \pi} ; i \lambda \right) - 1 \right] - 2 |\tau|
\label{torusgreens}
\eea
where we use the Jacobi Theta function
\beq
\theta_3(z;\tau) = \sum_{n = -\infty}^{\infty} \exp\left( i \pi \tau n^2 + 2 \pi i z n \right)
\eeq
For given values of $\tau$, $\varphi_x$, and $\varphi_y$, we compute the value of $L_y m_1$ by numerically inverting the gap equation 
$G_1(x,x;m_1^2) = 0$ using the Eq.~(\ref{torusgreens}). 

\bibliographystyle{apsrev4-1_custom}
\bibliography{entanglement}

\end{document}